\RequirePackage{vplref}[2005/04/25]
\documentclass[twocolumn,showpacs,aps,prd,superscriptaddress]{revtex4-1}
\usepackage{graphicx}
\usepackage{dcolumn}
\usepackage{epsfig}
\usepackage{amsmath}
\newcommand{\aone}{\ensuremath{a_1}}
\newcommand{\aonep}{\ensuremath{a_1^+}}

\newcommand{\BaBarYear}    {10}
\newcommand{\BaBarNumber}  {013}

\newcommand{\SLACPubNumber} {14195}

\newcommand{\BaBarType}      {PUB}  
\input babarsym
\RequirePackage{xspace}

\newcommand{\jhep}[1]{{\it JHEP}\ #1}

\newcommand{\calH}{\ensuremath{{\cal H}}}
\newcommand{\pvec}{{\bf p}}

\newcommand{\calB}{\ensuremath{{\cal B}}}

\newcommand{\bfemsix}{\ensuremath{\calB(10^{-6})}}

\newcommand{\DE}{\ensuremath{\Delta E}}

\newcommand{\xf}{\ensuremath{{\cal F}}}

\newcommand{\thetaT}{\ensuremath{\theta_{\rm T}}}
\newcommand{\costhr}{\ensuremath{\cos\thetaT}}

\newcommand\etal{{\it et al.}}
\newcommand{\half}{\ensuremath{\frac{1}{2}}}

\newcommand{\bfig}{\begin{figure}[htbpc!]}
\newcommand{\efig}{\end{figure}}
\newcommand\bef{\begin{figure}}
\newcommand\edf{\end{figure}}

\newcommand\sgline{\noalign{\vskip 0.10truecm\hrule\vskip 0.10truecm}}
\newcommand\beq{\begin{equation}}
\newcommand\eeq{\end{equation}}
\newcommand\bear{\begin{array}}
\newcommand\enar{\end{array}}
\newcommand\beqa{\begin{eqnarray}}
\newcommand\eeqa{\end{eqnarray}}
\newcommand\ben{\begin{enumerate}}
\newcommand\een{\end{enumerate}}

\newcommand{\UfourS}{\ensuremath{\Upsilon(4S)}}

\newcommand{\Kst}{\ensuremath{K^*}}

\newcommand{\Kstz}{\ensuremath{\Kstarz}}

   \newcommand{\rhop}{\ensuremath{\rho^+}}
   
   \newcommand{\rhoz}{\ensuremath{\rho^0}}

\newcommand{\bone}{\ensuremath{b_1}}

\newcommand{\BcacKstz}{\ensuremath{B^+ \to \aonep \Kstz}}
\newcommand{\rakst}{\ensuremath{0.7^{+0.5}_{-0.5} {}^{+0.6}_{-1.3}}}
\newcommand{\Rakst}{\ensuremath{(\rakst)\times 10^{-6}}}
\newcommand{\sakst}{\ensuremath{0.5}}
\newcommand{\ulakst}{\ensuremath{1.8}\xspace}
\newcommand{\Ulakst}{\ensuremath{\ulakst\times 10^{-6}}\xspace}
\newcommand{\ntotalpToy}{\ensuremath{15802}}
\newcommand{\resultY}{\ensuremath{61}}
\newcommand{\fitbias}{\ensuremath{34}}
\newcommand{\fitbiaserror}{\ensuremath{17}}
\newcommand{\rakstprodBR}{\ensuremath{1.3^{+1.1}_{-1.0} {}^{+1.1}_{-2.6}}}
\newcommand{\RakstprodBR}{\ensuremath{(\rakstprodBR)\times 10^{-6}}}
\newcommand{\ulakstprodBR}{\ensuremath{3.6}\xspace}
\newcommand{\UlakstprodBR}{\ensuremath{\ulakstprodBR\times 10^{-6}}\xspace}
\newcommand{\theTitle}{{\boldmath Search for $B^+$ meson decay 
to $a_1^+(1260)\Kstz$(892)}} 
\begin{document}
\begin{flushleft}
\babar-\BaBarType-\BaBarYear/\BaBarNumber \\
SLAC-PUB-\SLACPubNumber \\
\end{flushleft}

\title{\theTitle}

%
\author{P.~del~Amo~Sanchez}
\author{J.~P.~Lees}
\author{V.~Poireau}
\author{E.~Prencipe}
\author{V.~Tisserand}
\affiliation{Laboratoire d'Annecy-le-Vieux de Physique des Particules (LAPP), Universit\'e de Savoie, CNRS/IN2P3,  F-74941 Annecy-Le-Vieux, France}
\author{J.~Garra~Tico}
\author{E.~Grauges}
\affiliation{Universitat de Barcelona, Facultat de Fisica, Departament ECM, E-08028 Barcelona, Spain }
\author{M.~Martinelli$^{ab}$}
\author{A.~Palano$^{ab}$ }
\author{M.~Pappagallo$^{ab}$ }
\affiliation{INFN Sezione di Bari$^{a}$; Dipartimento di Fisica, Universit\`a di Bari$^{b}$, I-70126 Bari, Italy }
% ...
\collaboration{The \babar\ Collaboration}
\noaffiliation

\date{\today}

\begin{abstract}
We present a search for the decay
$B^+ \ra \aonep(1260) \Kstarz(892)$.
The data, collected with the \babar\ detector at the
SLAC National Accelerator Laboratory, represent 465 million \BB\ pairs
produced in \epem\ annihilation at the energy of the \FourS.
We find no significant signal and set an upper limit at 90\%\ confidence level 
on the product of branching fractions 
$\calB(B^+ \to \aonep(1260) \Kstz(892)) \times \calB(\aonep(1260) \to \pi^+ \pi^- \pi^+)$
of \Ulakst.
\end{abstract}


\maketitle


Measurements of the branching fractions and polarizations of charmless
hadronic \B\ decays are useful tests of the standard model and a means to
search for new physics effects.
In decays of \B\ mesons to a pair of
spin-one mesons, the longitudinal polarization, $f_L$, is 
particularly interesting. Simple helicity arguments favor 
$f_L$ to be close to 1, but several vector-vector ($VV$) 
decay modes such as $B\ra \phi\Kst$ \cite{phiKst}
and $B^{+}\ra\rhop\Kstz$ \cite{conjugate,rhopKst0},
are observed to favor $f_L \sim 0.5$. 
Possible explanations for this 
discrepancy have been proposed within the standard model 
\cite{VVSMrefs} as well as in new physics scenarios \cite{VVNPrefs}.

New ways to explore the size of contributing amplitudes in charmless
\B\ meson decays and their helicity structure may come
from measurements of the branching fractions and polarization of
charmless decays of $B$ mesons to an axial-vector meson and 
a vector meson ($AV$) or to an axial-vector meson and a 
pseudo-scalar meson ($AP$).
Theoretical decay rates have been predicted with the
na\"{i}ve factorization (NF) \cite{CMV} and QCD factorization (QCDF) 
\cite{C&Y} approaches.  The NF calculations find
the decay rates of $B\ra AV$ modes to be smaller than the corresponding
$\B \ra AP$ modes. The more complex QCDF calculations find the reverse.
For example, QCDF predicts a branching fraction of 
$(11^{+6.1}_{-4.4} {}^{+31.9}_{-9.0}) \times10^{-6}$ for $B^+ \to a_1^+ \Kstarz$ and 
$(32^{+16.5}_{-14.7} {}^{+12.0}_{-4.6}) \times10^{-6}$ 
for $B^0 \to b_1^- \rho^+$, while NF predicts a branching fraction of
$0.51 \times10^{-6}$ and $1.6 \times10^{-6}$, respectively.  
The first uncertainty on the QCDF prediction 
corresponds to the uncertainties due to the variation of Gegenbauer
moments, decay constants, quark masses, form factors and a \B\ meson wave function
parameter and the second uncertainty corresponds to the uncertainties due
to the variation of penguin annihilation parameters.
The NF prediction does not give an uncertainty on their value.

\B\ meson decays to charmless $AV$ final states are sensitive to penguin
annihilation contributions, which enhance some decay modes while suppressing others. 
Thus, investigating decays to many final states will help determine the size of
the contributing amplitudes.

A number of searches for $AV$ decays to the final states $a_1^{+}\rho^{-}$,
$\bone \rho$ and $\bone \Kst$ are presented in Ref. \cite{babar_a1rho}
and Ref. \cite{babar_b1V}, with upper limits on 
the branching fractions of $30\times 10^{-6}$ at 90\%\ confidence level (C.L.) for
$a_1^{+}\rho^{-}$ and from $1.4$ to $8.0\times 10^{-6}$ at 90\%\ C.L.
for the $\bone \rho$ and $\bone \Kst$ final states.
In this paper we present a search for the decay \BcacKstz.


The data for this measurement were collected with the \babar\
detector~\cite{BABARNIM} at the \pep2\ asymmetric-energy \epem\ 
storage ring
located at the SLAC National Accelerator Laboratory.  An integrated
luminosity of 424 \invfb, corresponding to $(465\pm5)\times 10^6$ \BB\
pairs, was produced in \epem\ annihilation at the $\Upsilon (4S)$
resonance (center-of-mass energy $\sqrt{s}=10.58\ \gev$).

A detailed Monte Carlo program (MC) is used to simulate the \B\
meson production and decay sequences, and the detector response
\cite{geant}. Dedicated samples of MC events for the decay
$B^+ \to \aonep \Kstarz$ with $\aonep \to \rho^0 \pi^+$
and $\Kstarz \to K^+ \pi^-$ were produced.
For the $\aonep$ meson parameters, we use 
the values given in Ref. \cite{PDG2008}
for studies with MC while 
for fits to the data we use a mass of $1229 \mevcc$ and a width of
$393 \mevcc$, which were extracted from 
$B^0 \to \aonep \pi^-$ decays \cite{BaBar_a1pi}.
We account for the uncertainties of these
resonance parameters in the determination of systematic uncertainties.
The $\aonep \to \pi^+\pi^-\pi^+$ decay
proceeds mainly through the intermediate states $\rho^0 \pi^+$ and 
$\sigma \pi^+$ \cite{PDG2008}. No attempt is made to separate
contributions of the dominant {\it P} wave $\rho^0$ from the
{\it S} wave $\sigma$ in the channel 
$\pi^+ \pi^-$. The difference in efficiency for the {\it S} wave and {\it P} wave cases is
accounted for as a systematic uncertainty.


We reconstruct \aonep\ candidates through the decay sequence
$\aonep\ra\rhoz\pip$ and $\rhoz\ra\pip\pim$. The other primary daughter
of the \B\ meson is reconstructed as $\Kstarz\ra\Kp\pim$. 
Candidates for the charged kaons must have 
particle identification signatures consistent with those of kaons.
Candidates for the charged pions must not be classified as protons, 
kaons, or electrons. 
We constrain the range of mass of reconstructed final-state candidates: 
between 0.55 and 1.0 \gevcc\ for the \rhoz, between 0.9 and 1.8
\gevcc\ for the \aonep, and between 0.8 and 1.0 \gevcc\ for the \Kstarz.

$B^+$ candidates are formed by combining \aonep\ and \Kstarz\ candidates.
The five final decay tracks in a candidate are fit to a common vertex. 
Candidates which have a $\chi^2$ probability for the fit greater than 0.01 are retained.
For these candidates, we calculate the energy substituted mass, 
$\mes=\sqrt{\frac{1}{4}s-\pvec_B^2}$, and the energy difference,
$\DE = E_B-\half\sqrt{s}$, where $(E_B,\pvec_B)$ is the \B\ meson 
energy-momentum four-vector, all values being expressed
in the \UfourS\ rest frame. We keep candidates with
$5.25\ \gevcc <\mes<5.29\ \gevcc$ and $|\DE|<100\ \mev$.  

We also impose restrictions on the helicity-frame decay
angle $\theta_{\Kstarz}$ of the \Kstarz\ mesons. 
The helicity frame of a meson is defined as the rest frame of that
meson, where the $z$ axis is the direction along which the
boost is performed from the parent's frame to this frame.
For the decay
$\Kstarz\ra K^+\pi^-$, $\theta_{\Kstarz}$ is the polar angle of the
daughter kaon, and for $\aonep\ra\rho^0\pi^+$, $\theta_{\aonep}$ is the polar
angle of the normal to the $\aonep \to 3\pi$ decay plane. 
We define $\calH_i = \cos(\theta_i)$, where $i=(\Kstarz,\aonep)$.
Since many background candidates accumulate near $|\calH_{\Kstarz}|=1$,
we require $-0.98 \le \calH_{\Kstarz} \le 0.8$.

Backgrounds arise primarily from random combinations of particles in
continuum $\epem\ra\qqbar$ events ($q=u,d,s,c$).  We reduce this 
background source with
a requirement on the angle \thetaT\ between the thrust axis
\cite{thrust} of the $B^+$ candidate in the \UfourS\ frame and that of the
charged tracks and neutral calorimeter clusters of the rest of the event.

The distribution is sharply peaked near $|\costhr|=1$
for jet-like continuum events, and nearly uniform for \B\ meson decays.
Optimizing the ratio of the signal yield to its (background dominated) 
uncertainty, we require $|\costhr|<0.8$.

A secondary source of background arises from $b \to c$ transitions. 
We reduce this background by eliminating events in which one of
the pions in the $B^+$ candidate is also part of a $D$ candidate. 

Such D candidates, reconstructed from $K^- \pi^+$ and $K^- \pi^+ \pi^+$, 
are required to have an invariant mass
within $0.02 \gevcc$ of the nominal $D$ meson mass.

The number of events which pass the selection is \ntotalpToy.
The average number of candidates found per event in the selected
data sample is 1.5 (2.0 to 2.4 in signal MC depending on the polarization).

We define a Neural Network (NN) for use in selecting the best
$B^+$ candidate. 
The $\chi^2$ probability of the vertex fit and the $\rho$ meson  mass
were the input variables to the NN.

To further discriminate against \qqbar\ background we construct a Fisher 
discriminant \xf\ \cite{fisher} which is a function of four variables: 
the polar angles of the $B^+$ candidate momentum and of the $B^+$ thrust axis
with respect to the beam axis in the \UfourS\ rest frame;
and the zeroth (second) angular moment $L_0$ ($L_2$) of the
energy flow, excluding the $B$
candidate, with respect to the $B$ thrust axis.
The moments are defined by $ L_j = \sum_i
p_i\times\left|\cos\theta_i\right|^j,$ where $\theta_i$ is the angle
with respect to the $B$ thrust axis of a track or neutral cluster $i$,
and $p_i$ its momentum.

We obtain yields and the longitudinal polarization $f_L$ from an extended
maximum  likelihood (ML)
fit with the seven input observables \DE, \mes, \xf, the resonance masses
$m_{\aonep}$ and $m_{\Kstarz}$, and the helicity 
variables $\calH_{\Kstarz}$ and $\calH_{\aonep}$.
Since the correlation between the observables in the selected data
and in MC signal events is small, we take the probability density function
(PDF) for each event to be a product of the PDFs for the individual
observables. Corrections for the effects of possible correlations, referred 
to as fit bias yield, are made on the basis of MC studies described below.
The components in the ML fit used are: signal, \qqbar\ background, 
charm \BB\ background, charmless \BB\ background, and
$B^+ \to a_2^+ \Kstarz$ background.

We determine the PDFs for the signal and \BB\ background components from
fits to MC samples. We develop PDF parameterizations for the combinatorial
background with fits to the data from which the signal region
($5.26\ \gevcc <\mes<5.29\ \gevcc $ and $|\DE|<60\ \mev$) has been excluded.

For the signal, the \mes\ and \DE\ distributions are parametrized
as a sum of a Crystal-Ball function~\cite{CB} and a Gaussian function.
In the case of \mes\ for \qqbar and \BB\
backgrounds we use the threshold function
$x\sqrt{1-x^2}\exp{\left[-\xi(1-x^2)\right]}$, where the argument
$x\equiv2\mes/\sqrt{s}$ and $\xi$ is a shape parameter. This function is 
discussed in more detail in Ref.~\cite{PRD04}. In the case of 
\DE\ for \qqbar and \BB\ backgrounds we use a polynomial function.
The PDFs for the Fisher discriminant ${\cal P}_j(\xf)$ are parametrized
as a single Gaussian function or a sum of two such functions.
The PDFs for the invariant masses of the \aonep\ and \Kstarz\ mesons 
for all components are constructed as sums of a relativistic Breit-Wigner 
function and a polynomial function.
We use a joint PDF ${\cal P}_j (\calH_{\Kstarz}, \calH_{\aonep})$  for the
helicity distributions. The signal and the $B^+ \to a_2^+ \Kstarz$ background 
component is parametrized as the product
of the corresponding ideal angular distribution in $\calH_{\Kstarz}$ and
$\calH_{\aonep}$ from Ref.~\cite{datta} times an empirical acceptance 
function ${\cal G}(\calH_{\Kstarz}, \calH_{\aonep})$, while the helicity PDF
for the other components is simply the product of the helicity PDFs for 
$\calH_{\Kstarz}$ and $\calH_{\aonep}$. The $\calH_i$ distributions for 
\qqbar and \BB\ backgrounds are based on Gaussian and 
polynomial functions. 

The likelihood function is
\begin{eqnarray*} 
  {\cal L} & = & \frac{e^{-{\left(\sum_j Y_j\right)}}}{N!}
  \prod_i^{N}\sum_j Y_j \times {\cal P}_j (\mes^i) {\cal P}_j(\xf^i) {\cal P}_j (\DE^i) \\
  & & {\cal P}_j (m_{\aonep}^i) {\cal P}_j (m_{\Kstarz}^i) {\cal P}_j
  (\calH_{\Kstarz}^i, \calH_{\aonep}^i), \label{eq:likelihood}
\nonumber  
\end{eqnarray*}
where $N$ is the number of events in the sample, and for each
component $j$ (signal, \qqbar\ background, $b\ra c$ transition \BB\ background,
charmless \BB\ background, or $B^+ \to a_2^+ \Kstarz$ background),
$Y_j$ is the yield of component $j$ and
${\cal P}_j(x^i)$ is the probability for variable $x$ of event $i$ 
to belong to component $j$. We allow the most important parameters
(first coefficient of the polynomial function for \DE, the invariant
masses of the \aonep\ and the \Kstarz, and
the width of the Breit-Wigner for the invariant mass of the \Kstarz) for
the determination of the combinatorial background PDFs to vary in the fit,
along with the yields for the signal, \qqbar\ background
and $b\ra c$ transition \BB\ background.

\begin{table}
\caption{Summary of results for $B^+ \to \aonep \Kstz$. Signal yield $Y$, 
    fit bias yield $Y_b$, the branching fraction
    $\calB = \calB(B^+ \to \aonep \Kstz) \times 
     \calB(\aonep \to \pi^+ \pi^- \pi^+)$,
    significance $S$ (see text) and upper limit UL. The 
    given uncertainties on fit yields are statistical only, while the
    uncertainties on the fit bias yield include the corresponding systematic
    uncertainties. The branching fraction of 
    $\Kstarz \to K^+ \pi^-$ is assumed to be $\frac{2}{3}$.}
\begin{center}
\begin{tabular}{cccccccc} \hline
$Y$ & $Y_b$ & $\bfemsix$ & $S$ & UL $(10^{-6})$ \\
\sgline
\hspace{1mm} $\resultY^{+23}_{-21}$ \hspace{1mm} & \hspace{1mm} $\fitbias \pm \fitbiaserror$ \hspace{1mm} 
& \hspace{1mm} \rakst \hspace{1mm} & \hspace{1mm} \sakst \hspace{1mm} & \hspace{1mm} \ulakst \hspace{1mm} \\
\hline
\end{tabular}
\end{center}
\label{tab:results}
\end{table}

We validate the fitting procedure by applying it to ensembles of
simulated experiments with the \qqbar\ component drawn from the PDF,
and with embedded known numbers of signal and \BB\ background
events randomly extracted from the fully simulated MC samples.  By
tuning the number of embedded events until the fit reproduces the yields
found in the data, we find a positive bias yield $Y_b$, to be subtracted
from the observed signal yield $Y$.
The fit bias yield arises from the neglected correlations
in signal and \BB\ background events.

 The corresponding numbers
are reported in Table~\ref{tab:results}.

We do not find a significant signal thus we do not report a 
measurement on the quantity $f_L$. In order to obtain the most
conservative upper limit, we assume $f_L=1$ in estimating the 
branching fraction. 

\begin{figure}
\begin{center}
\includegraphics[width=1.0\linewidth]{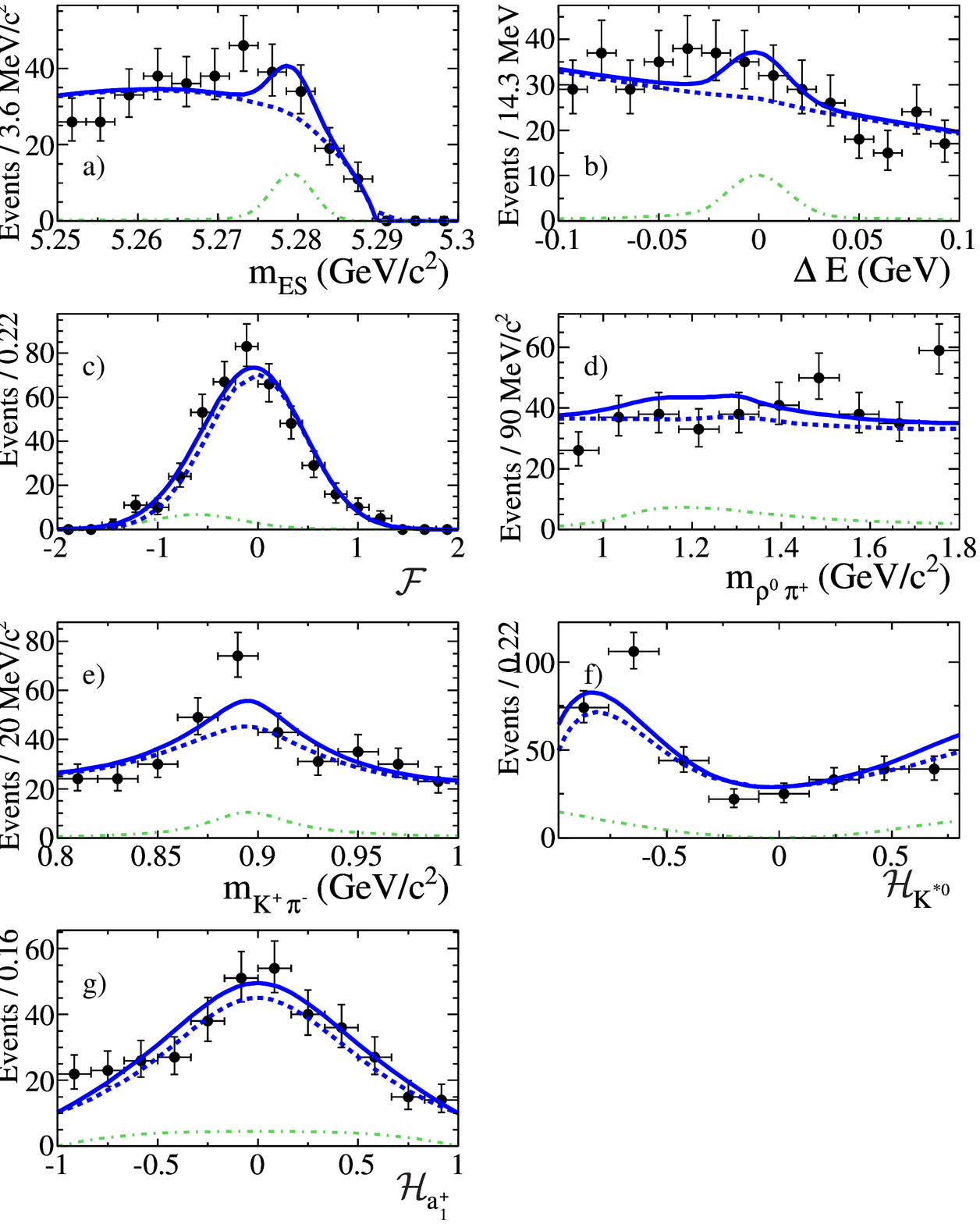}
\caption{Distributions for signal-enhanced subsets (see text)
  of the data projected onto the fit observables for the decay \BcacKstz;
  (a) \mes, (b) \DE, (c) \xf, (d) $m(\rho\pi)$ for the \aonep\
  candidate, (e) $m(K\pi)$ for the \Kstarz\ candidate, (f) $\calH_{\Kstarz}$
  and (g) $\calH_{\aonep}$.
  The solid lines represent the results of the fit, and the dot-dashed and
  dashed lines the signal and background contributions, respectively.
  These plots are made with requirements (see text) on the ratio of signal 
  to total likelihood, computed without the plotted variable. 
}
\label{fig:proj_all}
\end{center}
\end{figure}

We compute the branching fraction by subtracting the fit bias yield from the
measured yield and dividing the result by the number of produced \BB\
pairs and by the product of the selection efficiency and the branching
ratio for the ${\cal B}(\Kstarz\ra K^+ \pim)$ decay. We assume that the
branching fractions of the \UfourS\ to \BpBm\ and \BzBzb\ are equal,
consistent with measurements \cite{PDG2008}. The efficiency for 
longitudinally and transversely polarized signal events, obtained from the MC
signal model, is 12.9\% and 18.6\%, respectively. The results are given
in Table \ref{tab:results}, along with the significance, $S$, computed as the
square root of the difference between the value of $-2\ln{\cal L}$ 
(with additive systematic uncertainties included) for zero signal and
the value at its minimum. 
In Fig. \ref{fig:proj_all}\ we show the projections of data with PDFs
overlaid. The data plotted are subsamples enriched in signal with
the requirement of a minimum value of the ratio of signal to total
likelihood, computed without the plotted variable. We used 0.9 as 
requirement on the ratio in Fig. \ref{fig:proj_all}\ for each variable.
The efficiency of these requirements for signal is between $57\%$ and $70\%$
depending on the variable.


Systematic uncertainties on the branching fraction arise from the
imperfect knowledge of the PDFs, \BB\ backgrounds, fit bias yield, and 
efficiency.  
PDFs uncertainties not already accounted for by free parameters
in the fit are estimated from varying the signal-PDF parameters within 
their uncertainties. For \Kstarz\ resonance
parameters we use the uncertainties from Ref.~\cite{PDG2008} and for the
\aonep\ resonance parameters from Ref.~\cite{BaBar_a1pi}.
The uncertainty from fit bias yield (Table \ref{tab:results}) includes its
statistical uncertainty from the simulated experiments, and half of the
correction itself, added in quadrature.

To determine the systematic uncertainty arising from our imperfect
knowledge of the branching fractions of charmless \B\ decays, we vary
the charmless \BB\ background component yield by 100\%.
We conservatively assume that the branching ratio of $B^+ \to a_2^+ \Kstarz$
could be as large as that of $B^+ \to \aonep \Kstarz$ and vary the
$B^+ \to a_2^+ \Kstarz$ from 0 to 18 events around a fixed 
yield of 9 events used for the $B^+ \to a_2^+ \Kstarz$ component 
in the likelihood function.

The uncertainty associated with $f_L$ is estimated by taking the difference
in the measured branching fraction between the nominal fit ($f_L = 1$)
and the maximum and minimum values found in the scan along the
range [0, 1]. We divide these values by $\sqrt{3}$, motivated by
our assumption of a flat prior for $f_L$ in its physical range.

Uncertainties in our knowledge of the tracking efficiency are
0.24\% per track in the $B^+$ candidate.
The uncertainties in the efficiency from the event selection are
below 0.6\%.
The systematic uncertainty on the measurement of the integrated
luminosity is 1.1\%.
All systematic uncertainties on the branching fraction are summarized
in Table \ref{tab:sys}.

\begin{table}[!Bth]
\caption{Summary of systematic uncertainties of the determination of the 
         $B^+ \to \aonep \Kstz$ branching fraction.}
\begin{center}
\begin{tabular}{lr}
\hline
Source of systematic uncertainty       &  ~  \\
\sgline
Additive uncertainty (events)               &  ~  \\
~~~PDF parametrization                 &  4  \\   
~~~\aonep\ meson parametrization        &  6  \\
~~~ML fit bias yield                        &  \fitbiaserror \\
~~~Non resonant charmless \BB\ background        & 3  \\
~~~$B^+ \to a_2^+ \Kstarz$ charmless background  &  6  \\
~~~Remaining charmless \BB\ background &  7  \\
Total additive (events)                &  22 \\ 
\sgline
Multiplicative uncertainty (\%)                               & ~ \\ 
~~~Tracking efficiency                                   & 1.2 \\  
~~~Determination of the integrated luminosity            & 1.1 \\
~~~MC statistics (signal efficiency)                     & 0.6 \\
~~~Differences in selection efficiency for \aonep\ decay  & 3.3 \\  
~~~Particle identification (PID)                         & 1.4 \\  
~~~Event shape restriction ($\costhr$)                   & 1.0 \\  
Total multiplicative (\%)                                & 4.1 \\  
\sgline
Variation of $f_L$ \lbrack\bfemsix\rbrack                & ${}^{+0.0}_{-1.2}$ \\
\sgline
Total systematic uncertainty \lbrack\bfemsix\rbrack            & ${}^{+0.6}_{-1.3}$ \\
\hline
\end{tabular}
\end{center}
\label{tab:sys}
\end{table}


We obtain a central value for the product of branching fractions: 
\begin{eqnarray*}
\calB(B^+ \to \aonep \Kstz) \times \calB(\aonep \to \pi^+ \pi^- \pi^+)&~\\
           =\Rakst & \mathrm{,}& 
\end{eqnarray*}
where the first uncertainty quoted is statistical, the second systematic.
Including systematic uncertainties, this result corresponds to an 
upper limit at 90\%\ confidence level of \Ulakst.

Assuming
$\calB(\aone^{\pm}(1260) \to \pi^+ \pi^- \pi^{\pm} )$ is equal to
$\calB(\aone^{\pm}(1260) \to \pi^{\pm} \pi^0 \pi^0)$, and that
$\calB(\aone^{\pm}(1260) \to 3\pi)$ is equal to 100\%,
we obtain a central value:
\begin{eqnarray*}
\calB(B^+ \to \aonep \Kstz) = \RakstprodBR \mathrm{,}
\end{eqnarray*}
where the first uncertainty quoted is statistical, the second systematic.
Including systematic uncertainties, this result corresponds to an 
upper limit at 90\%\ confidence level of \UlakstprodBR.

This upper limit is in agreement with the prediction from na\"{i}ve 
factorization and lower than, but not inconsistent with, that of QCD 
factorization.

\par We are grateful for the 
extraordinary contributions of our \pep2\ colleagues in
achieving the excellent luminosity and machine conditions
that have made this work possible.
The success of this project also relies critically on the 
expertise and dedication of the computing organizations that 
support \babar.
The collaborating institutions wish to thank 
SLAC for its support and the kind hospitality extended to them. 
This work is supported by the
US Department of Energy
and National Science Foundation, the
Natural Sciences and Engineering Research Council (Canada),
the Commissariat \`a l'Energie Atomique and
Institut National de Physique Nucl\'eaire et de Physique des Particules
(France), the
Bundesministerium f\"ur Bildung und Forschung and
Deutsche Forschungsgemeinschaft
(Germany), the
Istituto Nazionale di Fisica Nucleare (Italy),
the Foundation for Fundamental Research on Matter (The Netherlands),
the Research Council of Norway, the
Ministry of Education and Science of the Russian Federation, 
Ministerio de Ciencia e Innovaci\'on (Spain), and the
Science and Technology Facilities Council (United Kingdom).
Individuals have received support from 
the Marie-Curie IEF program (European Union), the A. P. Sloan Foundation (USA) 
and the Binational Science Foundation (USA-Israel).


\renewcommand{\baselinestretch}{1}

\end{document}